# OptoGPT: A Foundation Model for Inverse Design in Optical Multilayer Thin Film Structures

Taigao Ma[1], Haozhu Wang[2†], L. Jay Guo[2*]

[1]Department of Physics, University of Michigan, Ann Arbor, Michigan 48109, USA

[2]Department of Electrical Engineering and Computer Science, University of Michigan, Ann Arbor, Michigan 48109, USA

[*]Email: guo@umich.edu

## Abstract

Optical multilayer thin film structures have been widely used in numerous photonic applications. However, existing inverse design methods have many drawbacks because they either fail to quickly adapt to different design targets, or are difficult to suit for different types of structures, e.g., designing for different materials at each layer. These methods also cannot accommodate versatile design situations under different angles and polarizations. In addition, how to benefit practical fabrications and manufacturing has not been extensively considered yet. In this work, we introduce OptoGPT (Opto Generative Pretrained Transformer), a decoder-only transformer, to solve all these drawbacks and issues simultaneously.

## Introduction

Optical multilayer thin film structure is one of the most vital photonic structures widely used in many applications, including structural color[1,2], filters[3], absorbers[4], distributed Bragg reflectors[5,6] (DBR), Fabry–Pérot[7] (FP) resonators, photovoltaic[8] and radiative cooling[9,10], among others. Inverse design seeks to identify the best material arrangements and obtain thickness combinations to achieve user-desired optical targets, which is critical to enable many of the above applications. Currently, there are two types of mainstream inverse design methods: 1) optimization-based methods[11–14], which rely on numerical simulations and iterative searches to minimize the difference between designed and targeted optical responses; and 2) deep learning-based methods[15–19], which use neural networks to learn a general mapping from the space of target responses to the space of optical multilayer thin film structures after training on a large dataset.

[†]: Work done while at University of Michigan. Currently at Amazon.

Although widely used, both methods have their own limitations, either from the perspective of design targets or types of designed structures. Optimization-based methods require running the algorithm from scratch when given a new or a different design target, which can be time-consuming. Deep learning-based methods are versatile for design targets, but existing works lack the ability to design for different types of structures (e.g., different material combinations at each layer; different total number of layers, etc). In addition, both methods seldomly examine how to expand the inverse design capabilities for angled incidence with different polarizations that are important for many applications, as well as simultaneous design under multiple conditions required for certain applications.

In addition to the above drawbacks, both methods also fail to accommodate the following two features that are vital for practical fabrications: diversity and flexibility. By diversity we mean that a single method can output multiple designs so that researchers can select for their fabrication based on the availability of materials and deposition methods, while flexibility allows researchers to arbitrarily impose restrictions on the material selection and thickness range at any layers for their fabrication or design needs. An inverse design method that can effectively meet these requirements will significantly bridge the gap between design and fabrication, making the design algorithm more practical.

In this work, we propose **OptoGPT** (Opto Generative Pretrained Transformer), a decoder-only transformer[20] model that can potentially address all these issues and unify the multilayer structure inverse design. To do so, first, we introduce "structure token" to fuse the representation of material and thickness and "structure serialization" to unify different types of structures. Next, we propose several techniques to unify the design target in different tasks as a combined reflection and transmission spectrum target. Further, a series of techniques based on "finetuning" and "probability sampling" are developed to unify the design under different angles and polarization, simultaneous design under multiple incident angles, as well as achieving diversity and flexibility for structure fabrication. Based on the empirical results demonstrated, we believe that OptoGPT can serve as a foundation model[21] for the design of optical multilayer thin films across a diverse array of applications.

## Methods

Designing a multilayer structure involves determining the material choice at each layer and the corresponding thicknesses of these layers. The major reason that existing deep learning-based methods cannot deal with different types of structures is that the output of these neural networks has fixed size that corresponds to a pre-defined structure, e.g., the three-layer structure of Ag/SiO$_2$/Ag in Ref. 16, the six-layer structure of MgF$_2$/SiO$_2$/Al$_2$O$_3$/TiO$_2$/Si/Ge in Ref. 19, and the twenty-layer structure of alternating SiO$_2$/Si$_3$N$_4$ in Ref. 15, etc. Therefore, these models can only design thickness for each layer and do not allow different material choices. This also make these models fail to accommodate structures with different number of layers. Here, we propose structure tokens and structure serialization to obtain the collaborative representation of materials and their thicknesses on the same footing, and treat the inverse design task as a conditional sequence generation problem.

### Structure Tokens and Structure Serialization

To address the aforementioned issues of existing approaches, we propose to treat material and thickness equally by concatenating them together to form a "structure token". Adding these tokens one by one, we can covert a multilayer structure into a sequence, which will be referred as "structure serialization". Figure. 1d gives one example of serializing a $N$-layer structure on the glass substrate using a sequence with $N + 1$ tokens. The first $N$ tokens describe the material and thickness at each layer and the last token is a special 'EoS' token that denotes the end of the sequence. Utilizing this approach, we can remove the limitation of fixed output size in the previous work and represent different types of structures (e.g., different material combinations at each layer; different total number of layers) in a unified approach.

In this work, we consider 18 different materials (see Figure. 1d; see Supplementary Information (SI) 1.1 for their refractive index data), and discretize the thickness in the range of [10, 500] nm with a step size of 10 nm. Therefore, for each layer, there are $18 \times 50 + 1 = 901$ possible tokens, corresponding to 900 different combinations of material and thickness plus one special 'EoS' token. We set the maximum number of layers to be 20, making the total number of multilayer structures under design consideration to be $(901)^{20} \sim 10^{59}$. We then need to search for an AI platform to accommodate such general approach and handle such a large sample space effectively.

**Conditional Sequence Generation**

Since the output format is now a sequence of tokens, the inverse design problem is equivalent to the sequence generation problem conditioned on the input of design targets. This is a problem that has been extensively researched and resolved in the Natural Language Processing (NLP) field using the Generative Pretrained Transformer[22,23] (GPT) model, especially the widely known ChatGPT[24]. Given some texts as input, e.g., a question or task description, GPT models can generate and output a text sequence that relates to the input. We propose to use similar GPT models to solve our problem. Differently, the input is the optical targets in the form of optical spectra (as a function of wavelength) while the output is the serialized physical structure of material and thickness. We show their analogy diagrams in Figure. 1a and Figure. 1b.

In this work, we set the reflection and transmission spectrum under normal incidence as our design target. The wavelength range covers the whole visible and near-infrared (NIR) region, spanning from 400 nm to 1100 nm with 10 nm step. We further propose a series of techniques that can expand the design target to absorption spectrum, and reflective/transmissive structural color with minimal adaptations (see below).

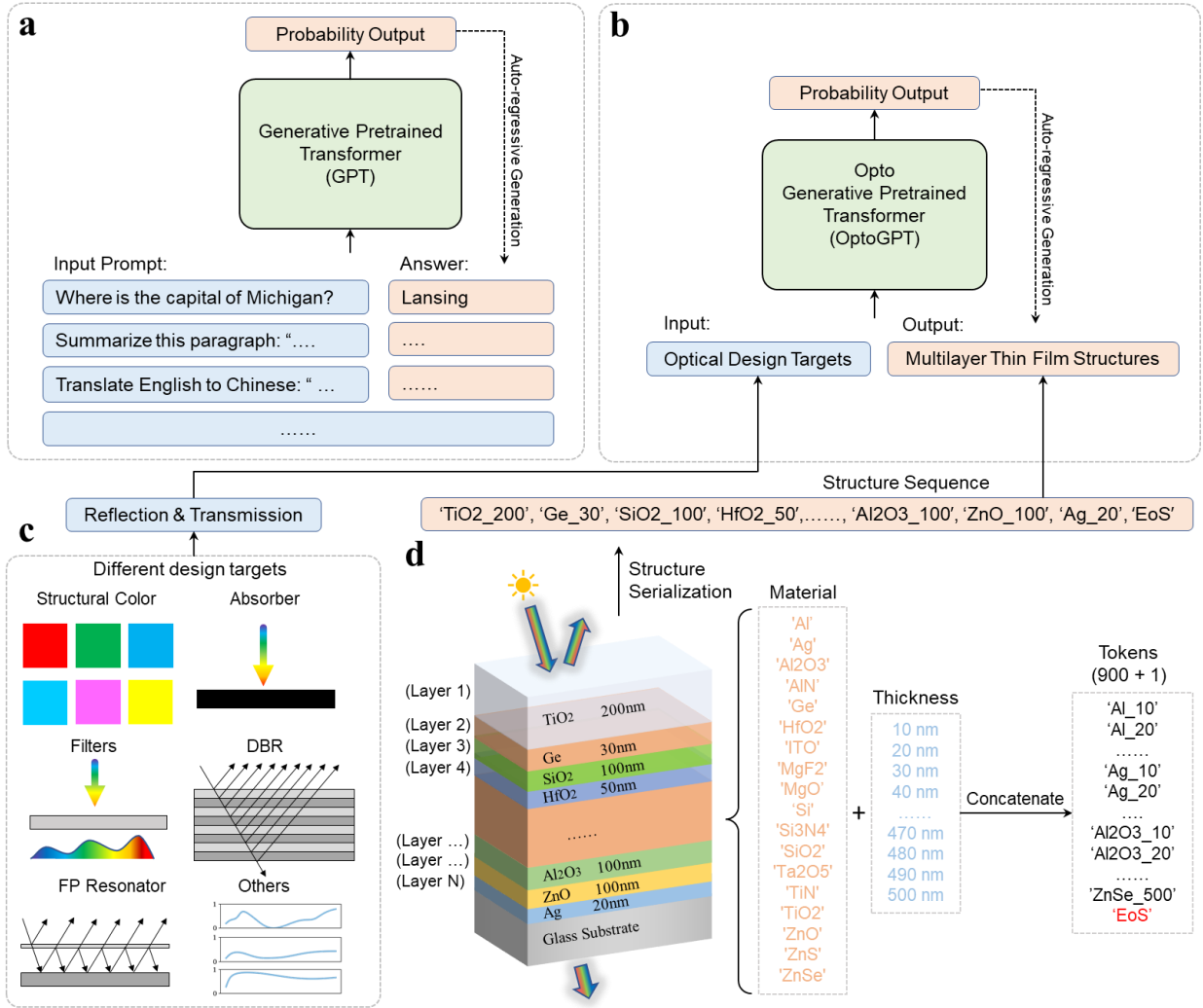

**Figure 1. The schematic of using the Opto-Generative Pretrained Transformer (OptoGPT) to design multilayer thin film structures.** (a) and (b) show the diagram of general GPT model in NLP and our OptoGPT, respectively. For GPT, the model takes in the input prompts and generate answers from probability sampling in an auto-regressive way. In OptoGPT, the input prompts are the optical targets while the outputs are designed multilayer structures. (c) Different types of inputs that relates to different application situations, including structural color, absorbers, filters, distributed bragg reflectors (DBR), Fabry–Pérot (FP) resonator and other arbitrary spectrum targets. All of them are converted to reflection and transmission spectrum. (d) One example of the "structure serialization" for a $N$-layer structure on the glass substrate. This $N$-layer structure is serialized by $N + 1$ tokens.

**Model Architecture**

Figure. 2a shows the architecture of our OptoGPT model. For the input, the spectrum target will go through a spectrum embedding to obtain its high-dimension hidden representation. For the output, the structure tokens will first go through a physical embedding layer to obtain its high-dimension hidden representation and then go through positional embeddings to obtain the relative position of each token inside this sequence. After that, both hidden representations of the input spectrum and output structures will go through a series of decoder blocks which contains attention layers, the major working mechanism behind GPT. The first self-attention layer is used to learn the relationship between layered structures, while the second cross-attention layer can capture the relationship between the input spectrum and the multilayer structure. Their output will further be used to give a probability distribution over all tokens. Our model is trained for ~200 epochs based on "next-word prediction"[20] using this probability output. We generate a large training dataset with 10 million samples and a validation dataset with 1 million samples (see SI 1.2). The total number of datasets is only $\sim 1/10^{52}$ of the possible structures. Each sample is a pair of a randomly sampled multilayer thin film structure on a glass substrate and the corresponding spectra simulated using Transfer Matrix Methods[25] (TMM). Details of training and model architecture can be found in SI 1.3. The model architecture and training details are summarized in SI 1.3, and visualization of multi-head self-attention is illustrated in SI 1.4.

**Inverse Design**

Once trained, our model can be used to design for a given input spectra target (see Figure. 2b), specifically, our model finishes the design layer-by-layer in an auto-regressive way (see Figure. 2c). When designing the $i_{th}$ layer, our model takes in the target spectrum together with the previously designed $i-1$ tokens, and outputs a probability distribution for all 900+1 tokens. Sampling from this distribution gives the design at the $i_{th}$ layer. These tokens will again be used as the input when designing the $(i+1)_{th}$ layer. This design process will keep going until reaching the maximum layer of 20 or 'EoS' is sampled.

This probability sampling has many advantages. First, because of the randomness during sampling, running each separate design process can output different structures. Therefore, the method inherently introduces diversity in the designed structure, capable of output multiple structures that satisfy the design target. In addition, it enables our model to design structures with different number of layers. For example, when 'EoS' is sampled at the fifth layer, our model terminates the design process and output the existing four-layer structure. The probability sampling will also be used to handle design with constraints in the latter section.

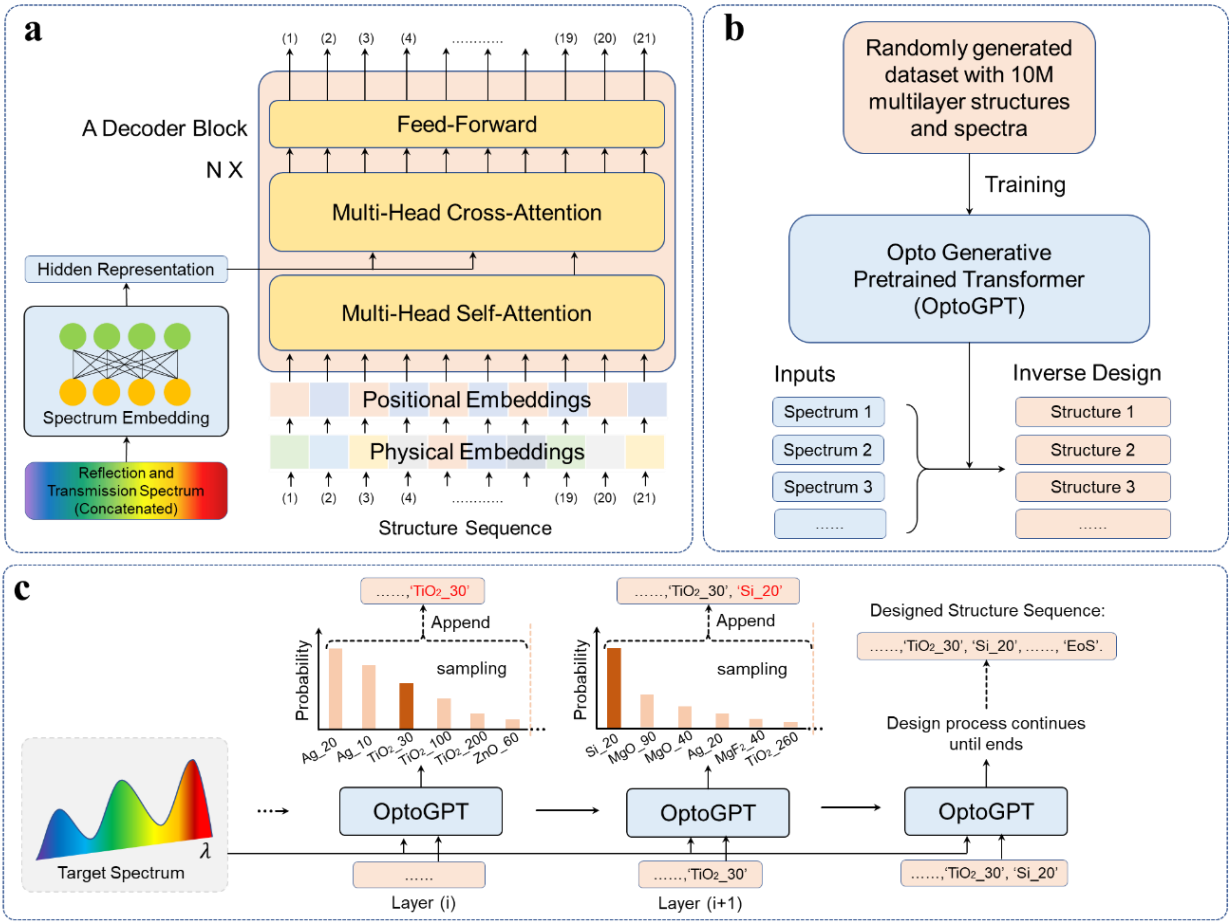

**Figure 2. Details of OptoGPT model.** (a) The model architecture of our OptoGPT, which is a decoder-only transformer. More details can be found in SI 1.3. (b) The working diagram of our OptoGPT model. (c) The diagram of the auto-regressive design process. When designing for $i_{th}$ layer, we sample from the probability output to obtain the layered information. This design process will keep going until reaching the maximum layer of 20 or 'EoS' is sampled.

## Results

**Visualization of Structure Tokens**

Before presenting the training and inverse design results, it is instructive to examine if the proposed structure tokens can capture the material and thickness information. We use the t-distributed stochastic neighbor embedding[26] (t-SNE) to reduce the dimension of their physical embeddings to 2-dimensions and visualize results in Figure. 3. To further compare these embeddings with the spectrum input, we also randomly select 1,000 spectra from the validation dataset and visualize their hidden representations' dimension reduction results in Figure. 3. Several interesting features are immediately observed. First, the physical structures (colored traces consisting of individual dots representing the structure token) and optical spectra responses (encircled cluster of green crosses) are well separated in this 2-D representation, even though they were fed into training on the equal footing. This demonstrates that our model has learned to distinguish the attributes of material structures and optical spectra while mapping them into the same hidden representation space.

Second, the 900 structure tokens are easily distinguishable, either as colored curves (the starting and ending points correspond to thickness of 500 nm and 10nm respectively), or cluster of dots, with no overlap between different materials. Upon close examination, it is clear that our model has intelligently separated the low refractive index dielectrics from the high refractive dielectrics (zoom-in view in (i) and (ii)). Within these two groups, all curves converge to the center region representing the lowest thickness 10 nm. This is anticipated from optical physics: when the dielectric layer thickness is reduced to the minimal all materials will behave similarly as they contribute to negligible optical phrase change or optical absorption (in the case of high index material). In other words, our model has learned the fact that thin dielectric layers of different materials all have similar effect on light propagation in multilayer thin films. Equally interesting is that all the metals cluster into their own territories in this 2-D map. This can be understood because as the metal layer thickness is greater than the optical penetration depth, its contribution to the optical response (i.e. spectra) has little dependence on the thicknesses. These observations demonstrate that even though our model does not directly take in any refractive index nor thickness, it can capture this information and learning hidden representations from a large dataset, validating the usage of structure serialization and spectrum embedding. This also aligns well with the strong representation capabilities demonstrated in many other foundation models such as Galactica[27], GaTo[28], and PaLM-E[29].

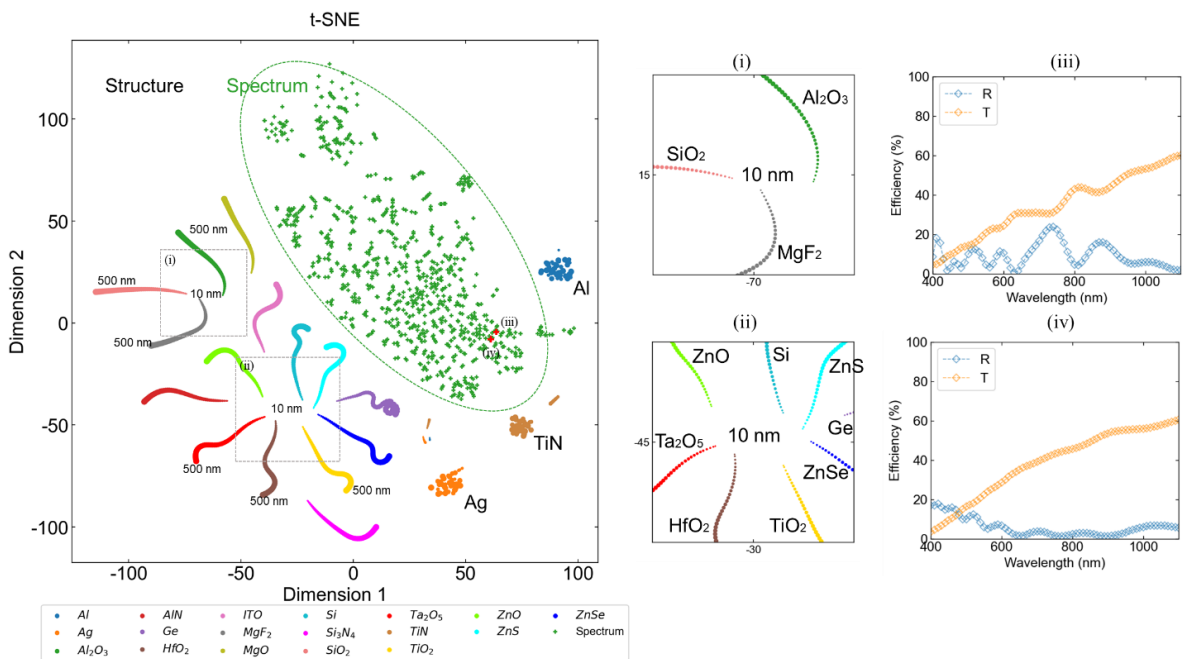

**Figure 3. 2D visualization of the hidden space using t-SNE to reduce dimension.** This figure includes 900 structure tokens and 1,000 spectra randomly selected from the validation dataset. Spectra are marked as green cross and structure tokens are marked as colorful dots, where different color corresponds to different materials. The green dashed circle illustrates the approximated boundary between spectra and structures. Inside this boundary are the spectra, with examples of two different spectra (marked as red cross) given in (iii) and (iv). Outside the green boundary are structure tokens corresponding to different material and thickness combinations. These structure tokens with the same materials either form a line shape or cluster together. For each line, the dot size is monotonically decreasing from one end to the other end, corresponding to the monotonical thickness decrease from 500 nm to 10 nm. Most lines converge into two regions, with zoom-in details given in (i) and (ii) corresponding to low refractive and high refractive index

region, respectively. Our model demonstrates the ability of learning the material and thickness from a large dataset without their explicit inputs.

**Inverse Design Performance**

Now we will examine our model's inverse design performance on different application situations. We want to mention that in this section, our model will be fixed and all these design tasks can be finished instantaneously by feeding different inputs of target optical response into our model. However, in case of higher accuracy is required, we run a thickness finetuning to improve the performance because the 10 nm discretization of thickness may lead to sub-optimal performance for certain materials (e.g. metals and absorbing dielectrics) (see SI 2.1). By default, we present the design performance without thickness finetuning unless specified.

*Performance on the Validation Dataset*

Here, we evaluate the averaged inverse design performance on 1,000 spectra targets randomly selected from the validation dataset. Based on the multilayer design output from our model, we simulate their corresponding spectrum using TMM and calculate the Mean Absolute Error (MAE) between the input spectrum and the simulated spectrum to quantify the design accuracy. The closest spectrum with smallest MAE in the training dataset is treated as the design baseline, i.e., the best spectrum we could get by simply referring to the training dataset. A good machine learning model should be able to learn from and outperform the training dataset. In Figure. 4a, we compare the MAEs of the closest structures in the training dataset (orange dots), designed structures (blue dots), and finetuned structures (red dots). On average, the MAE of the designed structures is 0.0258, which is lower than the MAE of the closest structures (0.0296) in the training set; finetuning the thickness can further reduce the MAE to 0.0192 (~24% reduction). In Figure. 4b, we compare the number of layers in the target structure (the structure corresponding to the target spectrum in the validation dataset) vs. the number of layers in the designed structure. The zero upper diagonal matrix implies that our model learns to solve design tasks using a simplified structure with fewer layers (~6 layers on average), which can facilitate the fabrication process as structures with fewer layers are easier to make. Finally, we record the time-consumption in Figure. 4c. On average, our model completes each design within 0.1s, which is comparable as running a TMM simulation. We detail one such inverse design example in Figure. 4d. By running the sampling process multiple times, our model can output multiple different structures with close-to-target spectrum that are much better than the training dataset. We also detail these designed structures to illustrate the diversity in Figure. 4d.

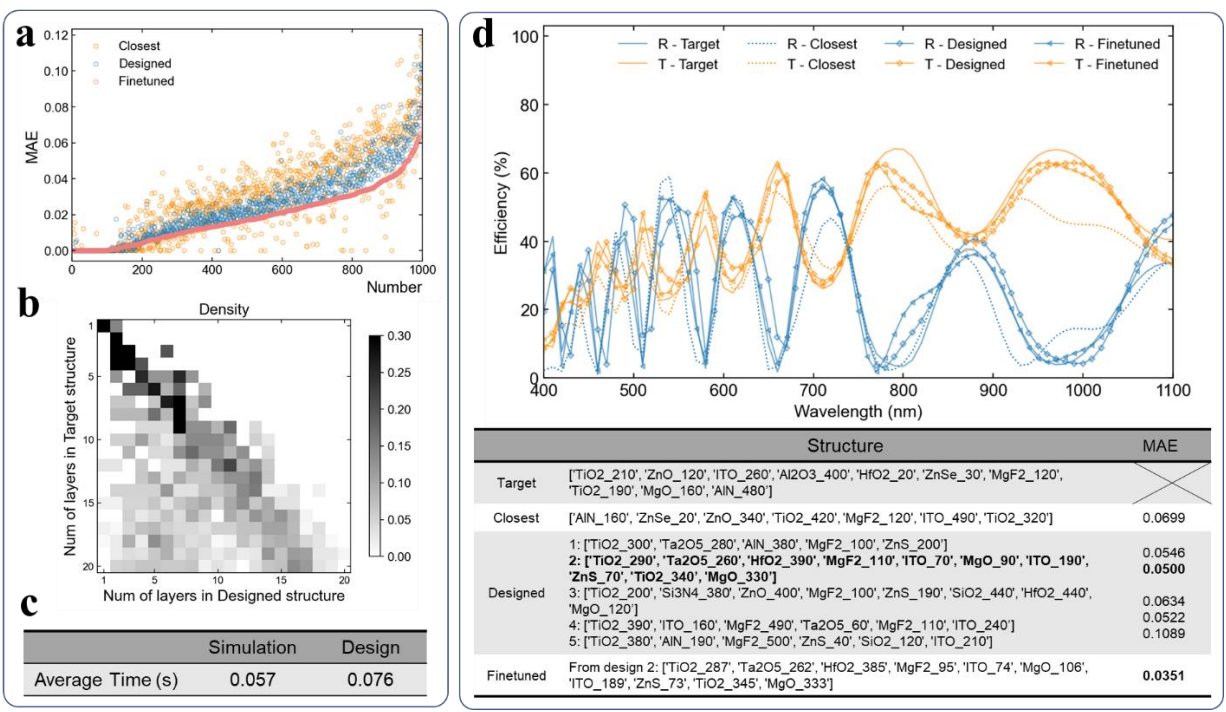

**Figure 4. Results of inverse design performance on the validation dataset**. (a) The Mean Absolute Error (MAE) on 1,000 random spectrum targets from the validation dataset. The orange, blue and red dots correspond to closest structures in training dataset, designed structures and finetuned structures. Their averaged MAE are 0.0296, 0.0258, 0.0192, respectively. (b) The number of layers in the target structure v.s. the number of layers in the designed structure. On average, the designed structures have 6 fewer layers than the target structure. (c) Time comparison of forward simulation using TMM and inverse design using OptoGPT (without finetuning). Results are averaged on 1,000 random targets in the validation dataset, showing that our model makes inverse design as fast as doing a TMM simulation. (d) One inverse design

example from the validation dataset. The table below gives the details of five designed structures and the finetuned structure as well as their spectrum MAE.

*Spectrum Filter*

Now we will evaluate our model on practical inverse design tasks. One such application is the spectrum filter which is used to selectively reflect or transmit specific band of light. Many deep learning-based methods have been proposed to inverse design these filters[15,30,31]. Here, several examples are tested: a band-notch filter at 550nm, a band-notch filter at 700nm, high reflection in NIR, double high reflection in 500-600nm and 800-1000nm, etc. We set the input to be the perfect rectangular spectrum, which has 0% transmission in the desired region and 100% transmission in the rest region. In all these artificial spectrum design targets, our model can output designs that outperform the training dataset. Thickness finetuning can further improve the accuracy. We illustrate two examples and compare their spectrum in Figure. 5a-b. More examples and details can be found in SI 2.5.

*Absorber*

Perfect absorbers have been widely used in photovoltaics, radiative cooling, detecting and solar-thermal harvesting[4,10,32–34], etc. Although our model is trained on reflection and transmission spectrum, it also demonstrates good performance for perfect absorbers. This is done by simply setting the input spectrum as zero for both reflection and transmission. Our model gives multiple designs and we show one design example in Figure. 5c. We also find that our model gives some similar structures reported in Ref. [35] and Ref. [19]. More details can be found in SI 2.4.

Apart from perfect absorbers, our model can also design for arbitrary absorption. Since energy conservation guarantees that reflection + transmission + absorption = 1, we can tailor the input spectrum by setting reflection to be one minus absorption and setting transmission to be zero. Figure. 5d gives one such design example. More design examples can be found in SI 2.4.

*Structural Color*

Compared to dyes and chemical pigments, structural colors[2] exhibit unique advantages on high resolution, stability and sustainability and have been widely used in color printing[36,37], information encryption[38], sensors[39], etc. There are also many works that use deep learning to solve the structural color inverse design[16,18,40]. Usually, colors can be represented by a three-dimensional color coordinate, e.g., Lab, RGB or xyY values. In order to make our model well suit for this application, here, we propose an algorithm that can convert a color coordinate into a continuous spectrum in a generalized way (see later section). These converted spectra can be pre-calculated and will not impact the design process. When designing for reflective color, we set the reflection spectrum to be this converted spectrum and transmission spectrum to be zero. For transmissive color, we set the transmission spectrum to be this converted spectrum and reflection spectrum to be one minus transmission. In Figure. 5e and Figure. 5f, we illustrate multiple design examples and their color visualization for both reflective and transmissive structural color, respectively. Details on the basic color theory, relationship of color coordinate and color spectra, and more examples can be found in SI 2.3.

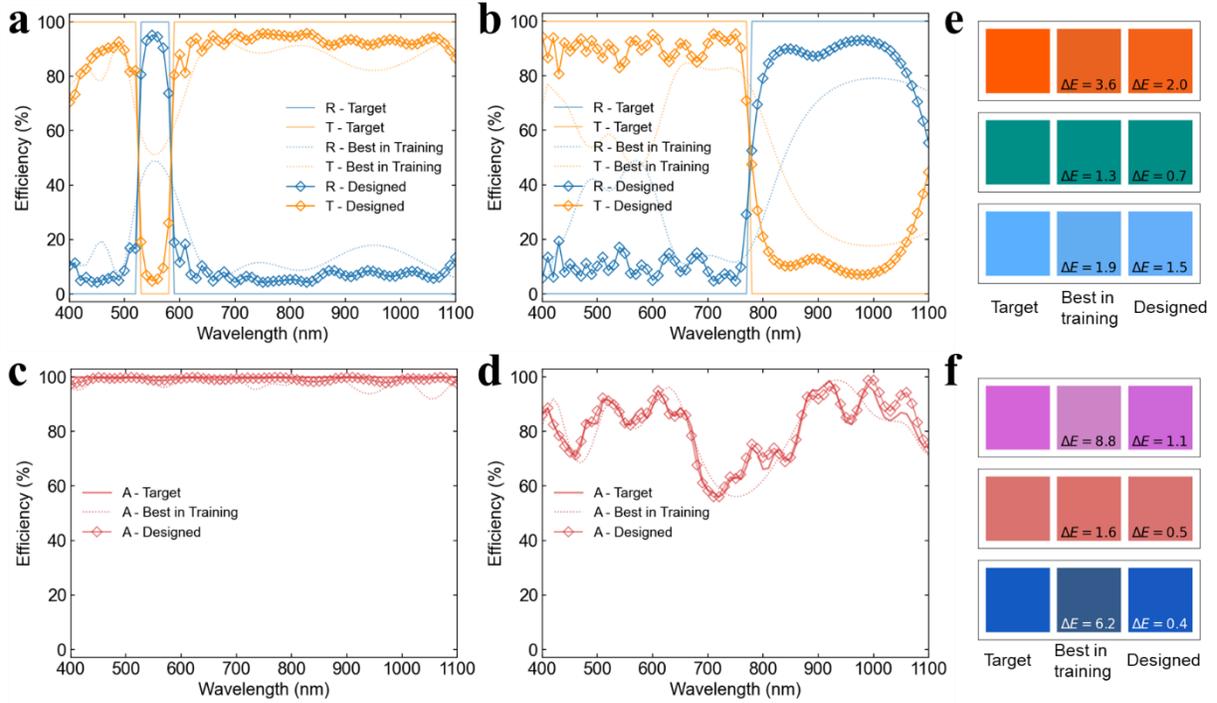

**Figure 5. Examples of inverse design artificial spectra in different applications.** (a) Design for band-notch filter at 550nm. (b) Design for high reflection in NIR. (c) Design for perfect absorber. (b) Design for arbitrary absorber. Here, real lines, dotted lines and squared lines correspond to the spectrum of artificial target, the spectrum of the closest structure in the training dataset and the spectrum of designed structure from our model (with thickness finetuning), respectively. (e-f) shows the example of designing reflective and transmissive structural color, respectively. We use the color difference of ΔE to evaluate the design performance (smaller ΔE means smaller color difference). For each color, the first brick, second brick, and third brick correspond to the target color, closest color in the training dataset, and designed color from our mode (with thickness finetuning), respectively. More details and examples can be found in section 2 in SI.

**Design Flexibility**

Design flexibility adds extra freedom to the design process because researchers can impose restrictions on the material selection and thickness range for any specific layer to meet the fabrication or design needs. We propose and apply a simple but generalized method of "probability resampling" to impose restrictions in the design process. As illustrated in Figure. 6a, this is done by removing these structures that do not satisfy constraints from probability sampling. Since this method is independent of spectra targets, it can be used to design for any applications. As an example, we use our model to inverse design a FP resonator. Here, the target spectrum has a resonance absorption at 610 nm and corresponds to a three-layer 20 nm Ag/150 nm SiO$_2$/50 nm Ag resonator on a glass substrate. We consider adding four different constraints separately:

1. Fix the first layer to be 100 nm SiO$_2$

2. Remove Ag in the third layer

3. Limit the thickness of the first layer within [10, 150] nm range and remove Ag/Al in the first layer

4. Specify the material arrangement to be a three-layer Ag/Si$_3$N$_4$/Ag structure and design the thickness only

The first constraint can be used when a dielectric layer at the air interface is needed for protection, while the second constraint is practical when looking for an alternative to replace silver, considering silver is an expensive metal. For the third constraint, we use it as a general example of adding thickness and material restrictions simultaneously. We use our model to design structures and we compare their spectra in Figure. 6b-d, demonstrating that our model can finish designs that satisfy desired constraints while still guaranteeing spectrum performance.

In particular, the fourth constraint specifies the material at each layer and only requires thickness design. This is a traditional design process widely used by human experts and in many optimization-based methods. The design results in Figure. 6e show that given the spectrum target and material arrangements, our model can be used for direct thickness design without iterative optimization. Since this feature does not rely on the target optical response, researchers can quickly examine if certain material combinations can achieve the target spectrum and obtain their corresponding thickness if so. We provide more examples of design flexibility in SI 3.2.

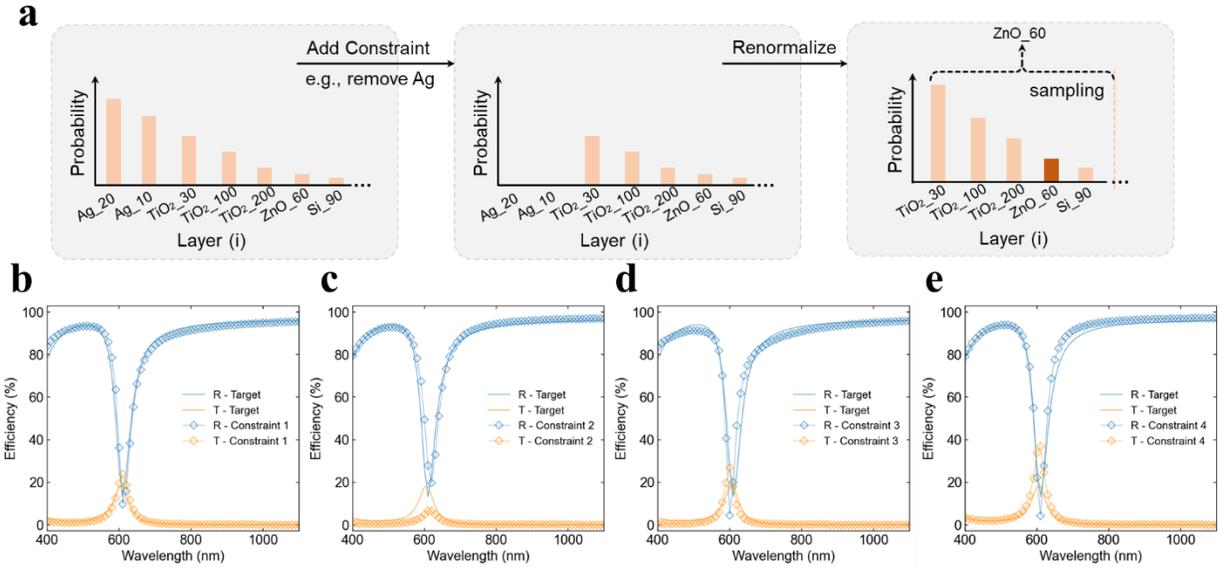

**Figure 6. Illustration of design flexibility.** (a) A visualization of the design process when adding the design constraint. We use the example of "remove Ag from material selection at $i_{th}$ layer". When designing the desired $i_{th}$ layer, we remove these tokens that do not satisfy constraints from probability distribution and only sample from the renormalized probability based on remaining tokens. (b-e) Comparison of the spectrum performance with different constraints, respectively. The real lines and squared lines are the target spectrum and the spectrum of the designed structure with different constraints, respectively. More examples of design flexibility can be found in section 3 in SI.

**Generalization Ability**

Although our model is trained on normal incident spectrum, its strong generalization ability enables the design towards different angles and polarization states, expanding allowable applications significantly. This is achieved through finetuning our entire model on a small dataset. We further propose the idea of "mixed sampling" to design structures that satisfy multiple requirements simultaneously.

*Finetuning*

Starting with the OptoGPT model trained on a 10M dataset, we can finetune this model on a smaller dataset to suit for light incidence of different angles and polarization states. Figure. 7a gives the finetune diagram. For example, in order to design for s-polarized spectrum under 20º incident angle, we first prepare a small 1M dataset with such spectrum and then update entire model by 10 epochs. This only requires 1% computing resources compared with training the entire model from scratch. Similar procedures can be done for other angles and polarizations. In Figure. 7b-g, we show multiple inverse design examples finetuned for 20° s-polarization, 60° s-polarization, 10° p-polarization, 50° p-polarization, 30° unpolarized light and 50° unpolarized light, respectively. More comparisons between pre-trained model and finetuned model are given in SI 4.1.

*Mixed Sampling*

Instead of designing the spectrum for a specific angle/polarization, in some situations we hope the designed structure can simultaneously realize multiple spectrum under different incident angles/polarizations, which has not been extensively explored. Benefited from our model's probability output, we can simply add up these outputs from multiple models that are specific to each situation, and then do a probability sampling based on this mixed output. This is called "Mixed Sampling" and we illustrate it in Figure. 8a. As an example, in Fig. 8b, we use this method to design an angle-invariant spectrum at 0°, 20° and 40° for unpolarized light. More examples can be found in SI 4.2.

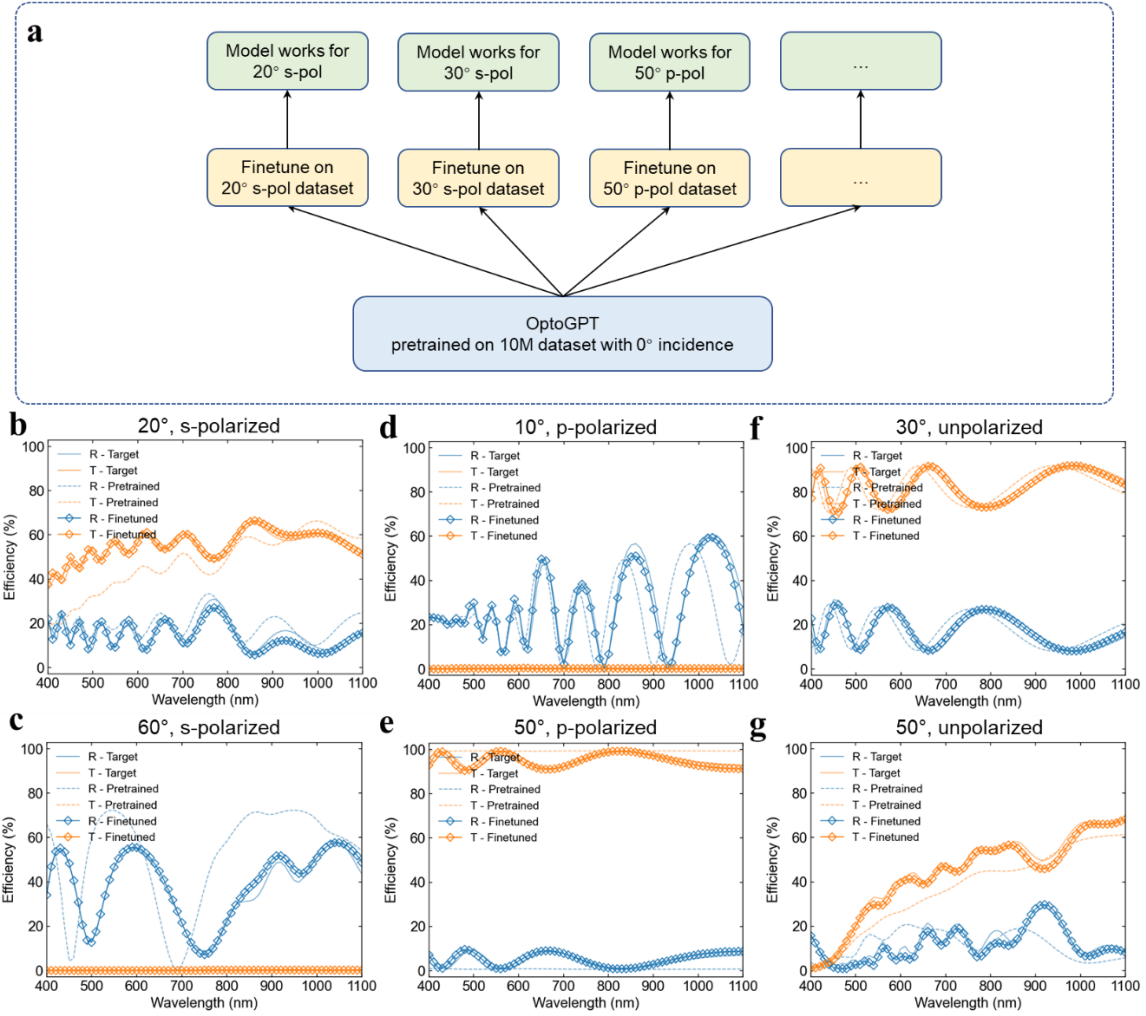

**Figure 7. Design performance on different angles and polarization.** (a) The diagram of finetune. (b-g) gives inverse design examples for spectrum with 20° s-polarization, 60° s-polarization, 10° p-polarization, 50° p-polarization, 30° unpolarized and 50° unpolarized, respectively. The real line, dashed line and squared line correspond to the target spectrum, spectrum designed by the pretrained model and spectrum designed by the finetuned model, respectively.

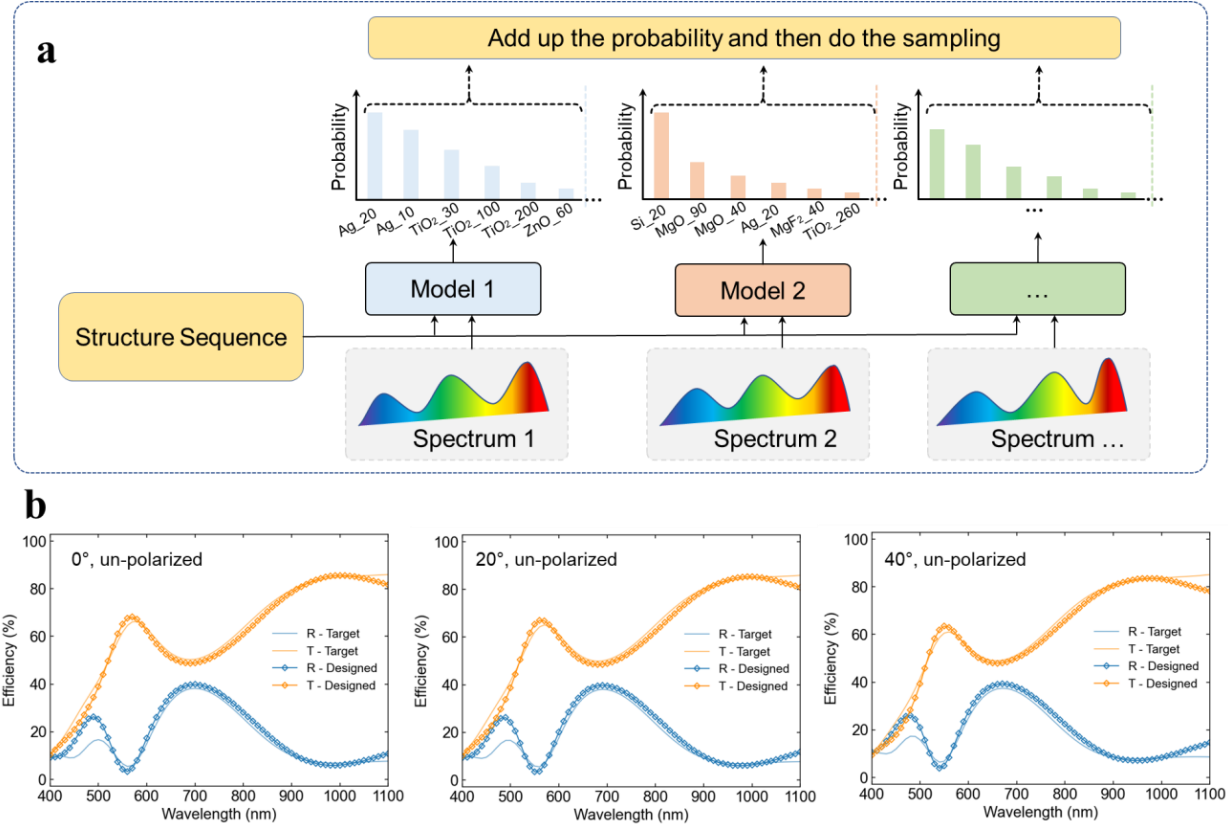

**Figure 8. Simultaneous design on different angles and polarization.** (a) The diagram of mixed sampling for simultaneous design. (b) One example of designing angle-robust spectrum for 0°, 20° and 40° unpolarized light.

## Discussion

By converting the multilayer structure into a sequence using structure tokens and structure serialization, we propose OptoGPT to effectively deal with the non-trivial inverse design problem in multilayer structure. Combined with many proposed techniques, our model can unify the inverse design under different types of input targets under different incident angle/polarization, be versatile to different types of structures, as well as facilitating the fabrication process by providing the diversity and flexibility. We compare our model with existing methods in Table 2 in SI. We hope the development of OptoGPT will make the multilayer thin film structure-based inverse design effective in methodology and easily accessible to researchers and engineers.

The interesting findings of the hidden representations of OptoGPT suggest that it has acquired domain-specific knowledge pertaining to optical multilayer structures through the training process. Furthermore, the model has demonstrated the capacity to apply this acquired knowledge effectively in the inverse design process. However, the current framework still lacks explain ability and does not allow users to directly understand the physical principles involved in its designs. For example, is there a general principle for designing absorbers and DBR? How to design high saturation structural color step-by-step? We hope future work can find a way to extract and formulate these design principles from the model and apply them to guide inverse design.

In addition, using similar methods, our model can be expanded towards high-dimension complicated photonic structures, e.g., 2D metasurfaces or 3D waveguides, using similar tokenization method in Vision Transformer[41]. However, one limitation is that our model requires a large dataset for training, which is also a common criticism for many GPT models. For example, ChatGPT is trained on billions of tokens using ~10,000 GPUs, costing ~$10M for a single training. In this work, because of the constraint on computation resources, we need to simplify our design problems, including using limited types of materials, limited spectrum range, thickness discretization as well as the maximum number of layers that can be designed, all of which can be extended with more computation resources. Despite using a large-scale dataset with 10 million samples for training, it is important to recognize that this dataset only covers a small fraction ($10^{-52}$) of the expansive and complex design space associated with optical multilayer thin film structures. Due to this limitation of its training dataset, OptoGPT may fail to find a design that lies outside the boundaries of the sampled design space (see SI 4.2). Close collaboration across multiple research groups is needed to obtain a better model for a more general and better photonic inverse design that expands to more complicated structures.

# References

1. Yang, Z., Ji, C., Liu, D. & Guo, L. J. Enhancing the Purity of Reflective Structural Colors with Ultrathin Bilayer Media as Effective Ideal Absorbers. *Advanced Optical Materials* **7**, 1900739 (2019).

2. Wang, D. *et al.* Structural color generation: from layered thin films to optical metasurfaces. *Nanophotonics* (2023) doi:10.1515/nanoph-2022-0063.

3. Ji, C. *et al.* Decorative near-infrared transmission filters featuring high-efficiency and angular-insensitivity employing 1D photonic crystals. *Nano Res.* **12**, 543–548 (2019).

4. Li, W., Xu, M., Xu, H.-X., Wang, X. & Huang, W. Metamaterial Absorbers: From Tunable Surface to Structural Transformation. *Advanced Materials* **34**, 2202509 (2022).

5. Hu, J. *et al.* Polariton Laser in the Bardeen-Cooper-Schrieffer Regime. *Phys. Rev. X* **11**, 011018 (2021).

6. Fink, Y. *et al.* A Dielectric Omnidirectional Reflector. *Science* **282**, 1679–1682 (1998).

7. Li, Z., Butun, S. & Aydin, K. Large-Area, Lithography-Free Super Absorbers and Color Filters at Visible Frequencies Using Ultrathin Metallic Films. *ACS Photonics* **2**, 183–188 (2015).

8. Liu, M., Johnston, M. B. & Snaith, H. J. Efficient planar heterojunction perovskite solar cells by vapour deposition. *Nature* **501**, 395–398 (2013).

9. Raman, A. P., Anoma, M. A., Zhu, L., Rephaeli, E. & Fan, S. Passive radiative cooling below ambient air temperature under direct sunlight. *Nature* **515**, 540–544 (2014).

10. Wang, S. *et al.* Scalable thermochromic smart windows with passive radiative cooling regulation. *Science* **374**, 1501–1504 (2021).

11. Rabady, R. I. & Ababneh, A. Global optimal design of optical multilayer thin-film filters using particle swarm optimization. *Optik* **125**, 548–553 (2014).

12. Tikhonravov, A. V., Trubetskov, M. K. & DeBell, G. W. Application of the needle optimization technique to the design of optical coatings. *Applied optics* **35**, 5493–5508 (1996).

13. Schubert, M. F. *et al.* Design of multilayer antireflection coatings made from co-sputtered and low-refractive-index materials by genetic algorithm. *Optics express* **16**, 5290–5298 (2008).

14. Shi, Y., Li, W., Raman, A. & Fan, S. Optimization of Multilayer Optical Films with a Memetic Algorithm and Mixed Integer Programming. *ACS Photonics* **5**, 684–691 (2018).

15. Liu, D., Tan, Y., Khoram, E. & Yu, Z. Training deep neural networks for the inverse design of nanophotonic structures. *Acs Photonics* **5**, 1365–1369 (2018).

16. Dai, P. *et al.* Inverse design of structural color: finding multiple solutions via conditional generative adversarial networks. *Nanophotonics* **11**, 3057–3069 (2022).

17. Unni, R., Yao, K. & Zheng, Y. Deep convolutional mixture density network for inverse design of layered photonic structures. *ACS Photonics* **7**, 2703–2712 (2020).

18. Wang, H. & Guo, L. J. NEUTRON: Neural Particle Swarm Optimization for Material-Aware Inverse Design of Structural Color. *iScience* 104339 (2022) doi:10.1016/j.isci.2022.104339.

19. Chen, W. *et al.* Broadband Solar Metamaterial Absorbers Empowered by Transformer-Based Deep Learning. *Advanced Science* **n/a**, 2206718 (2023).

20. Vaswani, A. *et al.* Attention is all you need. *Advances in neural information processing systems* **30**, (2017).

21. Bommasani, R. *et al.* On the Opportunities and Risks of Foundation Models. Preprint at http://arxiv.org/abs/2108.07258 (2022).

22. Radford, A., Narasimhan, K., Salimans, T. & Sutskever, I. Improving Language Understanding by Generative Pre-Training. 12 (2018).

23. Brown, T. B. *et al.* Language Models are Few-Shot Learners. Preprint at http://arxiv.org/abs/2005.14165 (2020).

24. Ouyang, L. *et al.* Training language models to follow instructions with human feedback. Preprint at http://arxiv.org/abs/2203.02155 (2022).

25. Byrnes, S. J. Multilayer optical calculations. *arXiv preprint arXiv:1603.02720* (2016).

26. Hinton, G. E. & Roweis, S. Stochastic Neighbor Embedding. in *Advances in Neural Information Processing Systems* vol. 15 (MIT Press, 2002).

27. Taylor, R. *et al.* Galactica: A Large Language Model for Science. Preprint at http://arxiv.org/abs/2211.09085 (2022).

28. Reed, S. *et al.* A Generalist Agent. Preprint at http://arxiv.org/abs/2205.06175 (2022).

29. Driess, D. *et al.* PaLM-E: An Embodied Multimodal Language Model. Preprint at http://arxiv.org/abs/2303.03378 (2023).

30. Han, X., Fan, Z., Liu, Z., Li, C. & Guo, L. J. Inverse design of metasurface optical filters using deep neural network with high degrees of freedom. *InfoMat* **3**, 432–442 (2021).

31. Unni, R., Yao, K., Han, X., Zhou, M. & Zheng, Y. A mixture-density-based tandem optimization network for on-demand inverse design of thin-film high reflectors. *Nanophotonics* **10**, 4057–4065 (2021).

32. Slobodkin, Y. *et al.* Massively degenerate coherent perfect absorber for arbitrary wavefronts. *Science* **377**, 995–998 (2022).

33. Lord, J. *et al.* Global potential for harvesting drinking water from air using solar energy. *Nature* **598**, 611–617 (2021).

34. Teperik, T. V. *et al.* Omnidirectional absorption in nanostructured metal surfaces. *Nature Photon* **2**, 299–301 (2008).

35. Yang, C. *et al.* Compact multilayer film structures for ultrabroadband, omnidirectional, and efficient absorption. *Acs Photonics* **3**, 590–596 (2016).

36. Tan, S. J. *et al.* Plasmonic Color Palettes for Photorealistic Printing with Aluminum Nanostructures. *Nano Lett.* **14**, 4023–4029 (2014).

37. Yang, W. *et al.* All-dielectric metasurface for high-performance structural color. *Nat Commun* **11**, 1864 (2020).

38. Song, M. *et al.* Color display and encryption with a plasmonic polarizing metamirror. *Nanophotonics* **7**, 323–331 (2018).

39. Balaur, E. *et al.* Colorimetric histology using plasmonically active microscope slides. *Nature* **598**, 65–71 (2021).

40. Gao, L., Li, X., Liu, D., Wang, L. & Yu, Z. A bidirectional deep neural network for accurate silicon color design. *Advanced Materials* **31**, 1905467 (2019).

41. Dosovitskiy, A. *et al.* An Image is Worth 16x16 Words: Transformers for Image Recognition at Scale. Preprint at http://arxiv.org/abs/2010.11929 (2021).

**Acknowledgements:**

We thank the National Science Foundation (PFI-008513 and FET-2309403) for the support of this work.

**Author Contribution:**

T. M carried out the research work, developed architecture and performed the data analysis. J. G conceived the research framework. H. W. provided constructive suggestions and feedback. The text was written by T. M., H. W., J. G.

**Competing Interests:**

The work reported in this manuscript is the subject of a U.S. patent application filed by the University of Michigan.